\documentclass[a4paper,fleqn,usenatbib]{mnras}

\usepackage{mathptmx}
\usepackage[T1]{fontenc}
\usepackage{ae,aecompl}
 
\usepackage{natbib}
\bibpunct{(}{)}{;}{a}{}{,}

\usepackage{amsmath}
\usepackage{amssymb}
\usepackage{graphicx}
\newcommand{\be}{\begin{equation}}
\newcommand{\ee}{\end{equation}}
\newcommand{\bea}{\begin{eqnarray}}
\newcommand{\eea}{\end{eqnarray}}

\newcommand{\m}{\,\hbox{m}}

\newcommand{\km}{\,\hbox{km}}

\newcommand{\cm}{\,\hbox{cm}}

\newcommand{\AU}{\,\hbox{AU}}
\newcommand{\g}{\,\hbox{g}}
\newcommand{\s}{\,\hbox{s}}

\newcommand{\Myr}{\,\hbox{Myr}}
\newcommand{\Gyr}{\,\hbox{Gyr}}

\title[Self-Stirring of Debris Discs]
{Self-Stirring of Debris Discs by Planetesimals\\ Formed by Pebble Concentration}

\author[Krivov \& Booth]{Alexander V. Krivov$^{1}$\thanks{E-mail: krivov@astro.uni-jena.de (AVK)}
and Mark Booth$^{1}$
\\
$^{1}$Astrophysikalisches Institut und Universit\"atssternwarte, Friedrich-Schiller-Universit\"at Jena,
      Schillerg\"a{\ss}chen~2--3, 07745 Jena, Germany
}

\date{Accepted {\em date}. Received {\em date}; in original form \today}

\pubyear{2018}

\begin{document}
\label{firstpage}
\pagerange{\pageref{firstpage}--\pageref{lastpage}}
\maketitle

\begin{abstract}
When a protoplanetary disc loses gas, it leaves behind planets
and one or more planetesimal belts.
The belts get dynamically excited, either by planets (``planet stirring'')
or by embedded big planetesimals (``self-stirring'').
Collisions between planetesimals become destructive and start to produce dust,
creating an observable debris disc.
Following Kenyon \& Bromley (2008), it is often assumed that self-stirring
starts to operate
as soon as the first $\sim 1000\km$-sized embedded ``Plutos'' have formed.
However, state-of-the-art pebble concentration models robustly predict planetesimals
between a few $\km$ and $\sim 200\km$ in size to form in protoplanetary discs rapidly,
before then slowly growing into Pluto-sized bodies.
We show that the timescale, on which these planetesimals excite the disc
sufficiently for fragmentation, is shorter than the formation timescale of Plutos.
Using an analytic model based on the Ida \& Makino (1993) theory,
we find the excitation timescale to be
$T_\text{excite} \approx 100 \; x_\text{m}^{-1} \; M_\star^{-3/2} \; a^3\Myr$,
where $x_\text{m}$ is the total mass of a protoplanetary disc progenitor
in the units of the Minimum-Mass Solar Nebula,
$a$ its radius in the units of $100\AU$, and
$M_\star$ is the stellar mass in solar masses.
These results are applied to a set of 23 debris discs that have
been well resolved with ALMA or SMA.
We find that the majority of these discs are consistent with being
self-stirred.  However, three large discs 
around young early-type stars do require planets as stirrers.
These are 49 Cet, HD~95086, and HR~8799,
of which the latter two are already known to have planets.
\end{abstract}

\begin{keywords}
planetary systems --
protoplanetary discs --
comets: general --
circumstellar matter --
submillimetre: planetary systems --
stars: individual: 49~Cet, HD~95086, HR~8799
\end{keywords}

%%%%%%%%%%%%%%%%%%%%%%%%%%%%%%%%%%%%%%%%%%%%%%%%%%

%%%%%%%%%%%%%%%%% BODY OF PAPER %%%%%%%%%%%%%%%%%%
 
\section{Introduction}

Debris discs around main-sequence stars are belts of planetesimals
that have not grown to full-size planets
\citep{wyatt-2008,krivov-2010,matthews-et-al-2013,hughes-et-al-2018,wyatt-2018}.
They are commonly observed through the thermal emission of the dust that these small bodies release in
collisions. To be able to produce that dust, planetesimal populations must be sufficiently stirred, i.e., 
planetesimals must have relative velocities 
high enough for collisions to be destructive.
Since the initial orbits of planetesimals formed in the protoplanetary phase are expected to be nearly 
circular and concentrate in the midplane of the disc, some mechanism is required to dynamically 
excite the planetesimal belts left after the gas dispersal.  Which stirring mechanism is at work is a 
matter of debate. Two main possibilities have been proposed: planetary stirring, i.e., stirring by planets 
orbiting the star interior or exterior to the planetesimal belt \citep{mustill-wyatt-2009} and 
self-stirring, i.e., excitation of small, field planetesimals
by big planetesimals (or dwarf planets) embedded in the belt 
\citep{kenyon-bromley-2010,kennedy-wyatt-2010}. 

This paper deals with the latter mechanism. The timescale on which the relative velocities of small 
planetesimals get sufficiently large for fragmentation, allowing them to start producing visible dust 
in collisions, is a sum of two timescales:
\be
  T_\text{stir} = T_\text{form} + T_\text{excite},
\label{eq:two timescales}
\ee
where 
$T_\text{form}$ is the formation timescale of big planetesimals and
$T_\text{excite}$ is the time it takes for these big stirrers to pump the random eccentricities 
and inclinations of field planetesimals to the values sufficient for fragmentation.

Formation of Pluto-sized bodies ($\sim 1000\km$ in radius) is normally considered sufficient
to trigger the cascade, although \citet{kenyon-bromley-2001} inferred that $500\km$-sized bodies
may suffice.
\citet{kenyon-bromley-2008} simulated the growth of big planetesimals in a disc, starting from km-sized seeds, 
in the runaway and oligarchic regimes, and found a convenient analytic formula for the timescale on which
the first ``Plutos'' emerge.
Assuming a standard Minimum-Mass Solar Nebula (MMSN)
with a solid surface density of $\sim 1 M_\oplus \AU^{-2} (r/\AU)^{-3/2}$ around a solar-mass star,
such objects would form on a timescale
\be
 T_\text{form}\sim T_\text{1000} \sim 400 (r/80\AU)^3\Myr.
\label{eq:1000 timescale}
\ee
Since bodies as big as $1000\km$ would excite the surrounding population of small planetesimals promptly,
the second term in Eq.~(\ref{eq:two timescales}) can be safely neglected, resulting in
\be
  T_\text{stir} \approx T_\text{form} .
\label{eq:one timescale}
\ee
This model suggests that young and large debris discs cannot be excited by self-stirring.
For instance, discs $110\AU$ in radius (which is close to the average radius of resolved debris discs,
see \citeauthor{pawellek-et-al-2014} \citeyear{pawellek-et-al-2014}) cannot be self-stirred in systems 
younger than $\approx 1\Gyr$.
This has been used to argue that as yet undiscovered planets must be responsible for triggering the 
collisional cascade in such systems, suggesting them as potential targets for planet searches
\citep[e.g.,][]{kennedy-wyatt-2010,moor-et-al-2014}.

However, this analysis comes with some caveats. First, formation of bodies smaller than $1000\km$ may 
already be sufficient to induce relative velocities above the fragmentation threshold, which would shorten
$T_\text{form}$. Second, Eq.~(\ref{eq:1000 timescale}) is only valid for a classical formation 
scenario, in which gravity-assisted collisions starting from km-sized seeds lead to slow, incremental 
growth of planetesimals
\citep[e.g.,][]{kenyon-luu-1999b,kenyon-bromley-2008,kobayashi-et-al-2010b,kobayashi-et-al-2016}. 
However, in recent years alternative formation pathways for planetesimals have been identified, 
most notably efficient local concentration of mm- to cm-sized ``pebbles''
in eddies of a turbulent disc
\citep{cuzzi-et-al-2008,cuzzi-et-al-2010,chambers-2010} or
by streaming instability
\citep[e.g.,][]{johansen-et-al-2015, simon-et-al-2016,carrera-et-al-2017, schaefer-et-al-2017, simon-et-al-2017}
with their subsequent gravitational clumping.
These models predict planetesimals with a spectrum of sizes 
between a few kilometres and a few hundred kilometres to emerge rapidly, on a few thousand dynamical 
timescales. If this scenario is at work, and the largest objects formed in this way are sufficient for
the required amount of stirring, then $T_\text{form}$ in Eq.~(\ref{eq:two timescales}) can be neglected, 
giving
\be
  T_\text{stir} \approx T_\text{excite} .
\label{eq:stir timescale}
\ee

The main goal of this paper is to investigate $T_\text{excite}$, i.e., the timescale on which bodies 
100s km in size, assumed to form ``instantaneously'', stir the surrounding population of small planetesimals, 
giving birth to a debris disc. We aim at a convenient formula for the pebble concentration scenario, 
similar to Eqs.~(\ref{eq:1000 timescale})--(\ref{eq:one timescale}) for the ``slow growth'' scenario.

Section~2 describes the analytic stirring model and numerical tests.
Section~3 gives the resulting formulae for the stirring timescales.
Section~4 applies the results to a sample of debris discs.
Section~5 contains a discussion.
Our conclusions are drawn in Section~6.

\section{Model}

Consider an annulus of radius $a$ and width $\Delta a$ populated by planetesimals.
We are interested in the stirring of smaller planetesimals by larger ones.
We address the problem in two steps.
In the first one, we consider two discrete populations:
small planetesimals of negligible mass and large ones of mass $M$.
In the second step, we generalize the model to a continuous mass distribution.

\subsection{Small field planetesimals and large stirrers}

To describe how planetesimals of mass $m$ are stirred by those of mass $M$,
we use the analytic theory of \citet{ida-1990} and \citet{ida-makino-1993}.
In this section, we assume small planetesimals to be test particles, i.e., set $m=0$.
In that case, both stirring of small planetesimals by themselves and dynamical 
friction are absent, and Eq.~(4.1) of \citet{ida-makino-1993} gives
\be
 {d\epsilon_\text{m} \over dt}
 = {\epsilon_\text{m} + \epsilon_\text{M} \over T \epsilon_\text{m}^2} ,
\label{eq:dx/dt}
\ee
where
$\epsilon_\text{m} = \left<e_\text{m}^2\right>$
are mean squares of orbital eccentricities of small planetesimals,
$\epsilon_\text{M} = \left<e_\text{M}^2\right>$
are those of large ones,
and $T  \epsilon_\text{m}^2$ is the viscous stirring timescale specified below.
We can assume the big planetesimals to move in circular orbits
($e_\text{M}=0$), so that Eq.~(\ref{eq:dx/dt}) simplifies to
\be
  {d\epsilon_\text{m} \over dt}
  = {T^{-1} \over \epsilon_\text{m}} .
\label{eq:dx/dt simple}
\ee

Assuming that the small planetesimals are in the dispersion-dominated regime,
the reciprocal of $T$ is given by Eq.~(4.2) of \citet{ida-makino-1993}:
\be
  T^{-1} =
  C_e \Omega a^2 
  \left( M \over M_\star \right)^2
  n_\text{M} ,
\label{eq:1/T orig}
\ee
where $C_e \approx 40$ is a numerical factor,
$\Omega$ is the mean motion,
$M_\star$ the mass of the central star,
and $n_\text{M}$ the surface number density of big planetesimals with mass $M$
in the ring of total mass $M_\text{disc}$:
\be
   n_\text{M} =
   {M_\text{disc} \over M}
   {1 \over 2\pi a \Delta a} ,
\ee
so that
\be
  T^{-1} =
  {1 \over 2 \pi}
  C_e \Omega
  \left( a \over \Delta  a\right)
  \left( M \over M_\star \right)
  \left( M_\text{disc} \over M_\star \right).
\label{eq:1/T}
\ee

The solution to Eq.~(\ref{eq:dx/dt simple}) is
\be
 \sqrt{\left<e_\text{m}^2\right>}
 =
 \left( 2t/T \right)^{1/4}.
\label{eq:e(t)}
\ee
Similar equations hold for the orbital inclinations: 
$\epsilon_\text{m} = \left<e_\text{m}^2\right>$ and 
$\epsilon_\text{M} = \left<e_\text{M}^2\right>$ 
are replaced by
$\iota_\text{m} = \left<I_\text{m}^2\right>$ and 
$\iota_\text{M} = \left<I_\text{M}^2\right>$,
whereas the factor $C_e$ is replaced by $C_I \sim 2$.
Since $C_I \ll C_e$, the eccentricities grow much faster than the inclinations,
so that the contribution of the inclinations to the relative velocities between small 
planetesimals can be neglected:
\be
  v_\text{rel} \approx v_\text{k} \sqrt{\left<e_\text{m}^2\right>} ,
\ee
where $v_\text{k}$ is the circular Keplerian speed at a distance $a$ from the star.

\begin{figure}
\centering
% Original at /astro/krivov/SMALL_PROJECTS/SELF_STIRRING/ANAL/ecc_evolution.pdf
\includegraphics[width=0.49\textwidth, angle=0]{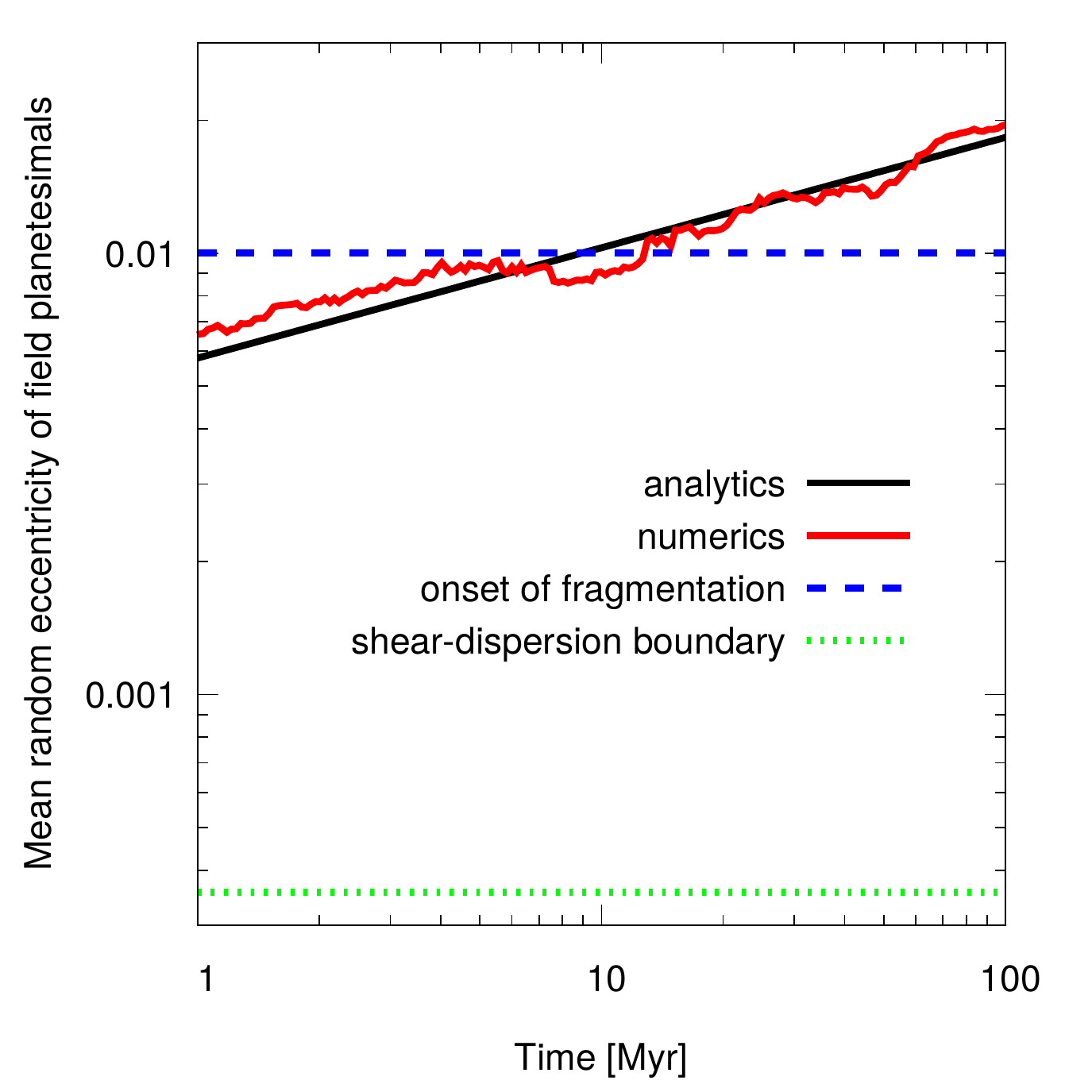}
\caption{
Time evolution of the root mean square eccentricity
of the field planetesimals in the reference case
($M_\star = M_\odot$,
$a = 100\AU$, $\Delta a = 10\AU$,
$M_\text{disc} = 100 M_\oplus$,
$M = 5.8 \times 10^{-6} M_\oplus$).
Black solid line:  Eqs.~(\ref{eq:1/T})--(\ref{eq:e(t)}).
Red solid line: scaled Mercury6 run, as described in the text.
Horizontal lines are characteristic eccentricity levels:
an eccentricity corresponding to the boundary between shear-dominated and dispersion-dominated 
regimes (green dotted) and
an eccentricity corresponding to the onset of fragmentation, i.e., to the relative velocities of 
$30\m\s^{-1}$ (blue dashed).
\label{fig:e(t)}
   }
\end{figure}

If the mean relative velocity of small planetesimals increases to a certain value $v_\text{frag}$
sufficient for fragmentation, the disc gets sufficiently stirred to trigger the collisional cascade
and to become a debris disc.
Denoting
\be
  e_\text{frag} =  v_\text{frag} / v_\text{k},
\ee
the stirring timescale amounts to
\be
  T_\text{stir} =  {T \over 2} e_\text{frag}^4 .
\label{eq:Tstir efrag}
\ee

The above theory is only valid if the feeding zones of the stirrers overlap.
This implies a condition on $M_\text{disc}$, $M$, and the relative ring width $\Delta a /a$ that is easy to derive.
The width of the feeding zone can be roughly estimated as \citep[][their Eq. 3.4]{ida-makino-1993}
\be
 \Delta a_\text{f}  \approx 8\sqrt{3} h_\text{M} a_\text{M}
\ee
with
\be
 h_\text{M} = \left( M/(3M_\star) \right)^{1/3} .
\ee
The feeding zones overlap if
\be
 N \Delta a_\text{f} \ge \Delta a ,
\label{eq:overlap1}
\ee
where $N = M_\text{disc}/M$ is the number of stirrers.
This yields the desired condition:
\be
 {\Delta a \over a}
 \le
 C
 \left( M_\text{disc} \over M_\star \right)^{1/3}
 \left( M_\text{disc} \over M       \right)^{2/3} 
\label{eq:overlap2}
\ee
with $C\equiv 8 \times 3^{1/6} \approx 10$.

To illustrate how the model works, we chose the following ``reference case''.
We assumed a solar-mass central 
star ($M_\star = M_\odot$),
the ring of small planetesimals of radius $a = 100\AU$
and width $\Delta a = 10\AU$,
having the mass $M_\text{disc} = 100 M_\oplus$ ($M_\oplus$ is the Earth mass),
and large stirrers of mass $M = 3.4 \times 10^{22}\g$ (or $5.8 \times 10^{-6} M_\oplus$;
corresponding to the radius of big planetesimals
of $200\km$ for the bulk density of $\rho = 1.0 \g\cm^{-3}$).
For this setup, condition (\ref{eq:overlap2}) is fulfilled.

The black solid line in Fig.~\ref{fig:e(t)} depicts the time evolution of 
$\sqrt{\left<e_\text{m}^2\right>}$ in this reference case, as predicted by 
Eqs.~(\ref{eq:1/T})--(\ref{eq:e(t)}).
We also show two typical eccentricity values with horizontal straight lines.
The lowest line at $\sqrt{\left<e_\text{m}^2\right>} = 2 h_\text{M}$ is a 
boundary between the shear-dominated and dispersion-dominated regimes \citep{ida-makino-1993}.
The uppermost line is the rms eccentricity that corresponds to the relative velocities
of $v_\text{frag}=30\m\s^{-1}$.
This is roughly the minimum impact velocity needed to disrupt kilometre-sized planetesimals
kept together by gravity \citep[e.g.,][]{benz-asphaug-1999}.
As soon as this line is crossed, which happens in $\lesssim 10\Myr$ in the reference case,
we consider the disc to be sufficiently stirred for fragmentation to occur.

We tested the analytic model with numerical integrations,
using the Mercury6 package with the Bulirsch-Stoer integrator
\citep{chambers-1999}.
The setup was the same as in Sect.~III.2 of \citet{ida-makino-1993}.
As in their runs, we took a central star of mass $M_\star = M_\odot$,
a ring of radius $a = 1\AU$ and width $\Delta a = 2 \times 17.4 h_\text{M} a = 0.11\AU$,
and one stirrer with $M = 0.035 M_\oplus$.
That stirrer was placed in a nearly circular, non-inclined orbit with 
$e_\text{M} = I_\text{M}= 0.01 h_\text{M} = 3.2 \times 10^{-5}$.
As \citet{ida-makino-1993}, we traced 800 field planetesimals, each having
the mass of $0.01M$.
We were able to closely reproduce their results (Figs.~3 to~5 in their paper).
We then re-scaled the timescale of this setup to the reference one
by means of Eq.~(\ref{eq:1/T}),
multiplying the timescale of that test by a factor of
\be
 \left( a^\text{ref} \over a^\text{IM} \right)^{3/2}
 \left( \Delta a^\text{ref}/ a^\text{ref}
        \over  
        \Delta a^\text{IM}/ a^\text{IM}
 \right)
  \left( M^\text{ref} \over M^\text{IM} \right)^{-1}
  \left( M_\text{disc}^\text{ref} \over  M_\text{disc}^\text{IM} \right)^{-1}
 \approx
 3200 ,
\ee
where the superscripts ``ref'' and ``IM'' stand for the parameter values of the
reference case and the Ida \& Makino numerical setup, respectively.
The numerical result is overplotted in Fig.~\ref{fig:e(t)} with a red solid line.
A comparison with the analytic curve demonstrates a reasonable match between the numerics
and analytics and the validity of scalings.
Besides, the numerical result proves that the field planetesimals get into the 
dispersion-dominated regime pretty quickly, in $\ll 1\Myr$, even if they start from
initially circular orbits.
This ensures that Eq.~(\ref{eq:1/T orig}) is valid.

Since the \citet{ida-makino-1993} setup described above is very far from
the configurations typical of debris discs we are interested in here,
we performed a few additional Mercury6 runs.
In those runs, we varied the orbital radius $a$ of the perturber and its mass $M$.
We also tried setting the mass of small planetesimals to zero.
Again, the results of each run were re-scaled to the reference case
with the aid of Eq.~(\ref{eq:1/T}).
In all the cases the analytics and numerics agreed to each other within a factor of two,
which we deem sufficient for our purposes.

\subsection{Planetesimals with a mass distribution}

We now assume planetesimals in the ring to have a power-law size distribution
from some minimum radius $s$ (or mass $m$)
to a maximum radius $S_\text{max}$ (or mass $M_\text{max}$).
The exact values of $s$ (or $m$) are unimportant: it is only required that
$s \ll S_\text{max}$ (or $m \ll M_\text{max}$).

Denoting by $\epsilon(m)$ the mean squares of orbital eccentricities of planetesimals with mass $m$,
by $n(m)dm$ the surface number density of planetesimals with masses in $[m, m+dm]$,
Eqs.~(\ref{eq:dx/dt}) and (\ref{eq:1/T orig}) generalise to
\be
 {d\epsilon(m) \over dt}
 =
  C_e \Omega a^2 
  \int_m^{M_\text{max}}
  \left( M \over M_\star \right)^2
  n(M)
{\epsilon(m) + \epsilon(M) \over \epsilon^2(m)}
  dM .
\label{eq:dx/dt cont}
\ee
To keep the problem solvable analytically, we choose to eliminate $\epsilon(M)$ from the integrand.
Assuming that $\epsilon(M) \le \epsilon(m)$, which is natural to expect from the dynamical friction,
there are two obvious possibilities to do that.
The first is to set $\epsilon(M)=0$, i.e., to neglect the eccentricity
of the stirrers (which are all bodies with mass $>m$). This would result in a slower growth of $\epsilon(m)$ 
than it actually is, and thus in an upper limit on the stirring timescale. The second one is to set
$\epsilon(M) = \epsilon(m)$, i.e., to assume that all the bodies are in orbits with the same eccentricities, 
regardless of their mass. This would have the opposite effect, overpredicting the growth rate of 
$\epsilon(m)$ and resulting in the lower limit on the stirring timescale.
The exact solution would be between these two limiting cases.
Thus we rewrite Eq.~(\ref{eq:dx/dt cont}) as
\be
  {d\epsilon(m) \over dt}
  = \gamma {T^{-1}(m) \over \epsilon(m)} .
\label{eq:dx/dt cont simple}
\ee
with
\be
 T^{-1} (m)
 =
  C_e \Omega a^2 
  \int_m^{M_\text{max}}
  \left( M \over M_\star \right)^2
  n(M)
  dM ,
\label{eq:1/T cont orig}
\ee
where $1 \le \gamma \le 2$.
Eqs.~(\ref{eq:dx/dt cont simple}) and (\ref{eq:1/T cont orig}) generalise
Eqs.~(\ref{eq:dx/dt simple}) and (\ref{eq:1/T orig}), respectively, to the continuous distribution
of planetesimals.

\begin{figure*}
\centering
% Original at /astro/krivov/SMALL_PROJECTS/SELF_STIRRING/ANAL/timescales.pdf
\includegraphics[width=1.0\textwidth, angle=0]{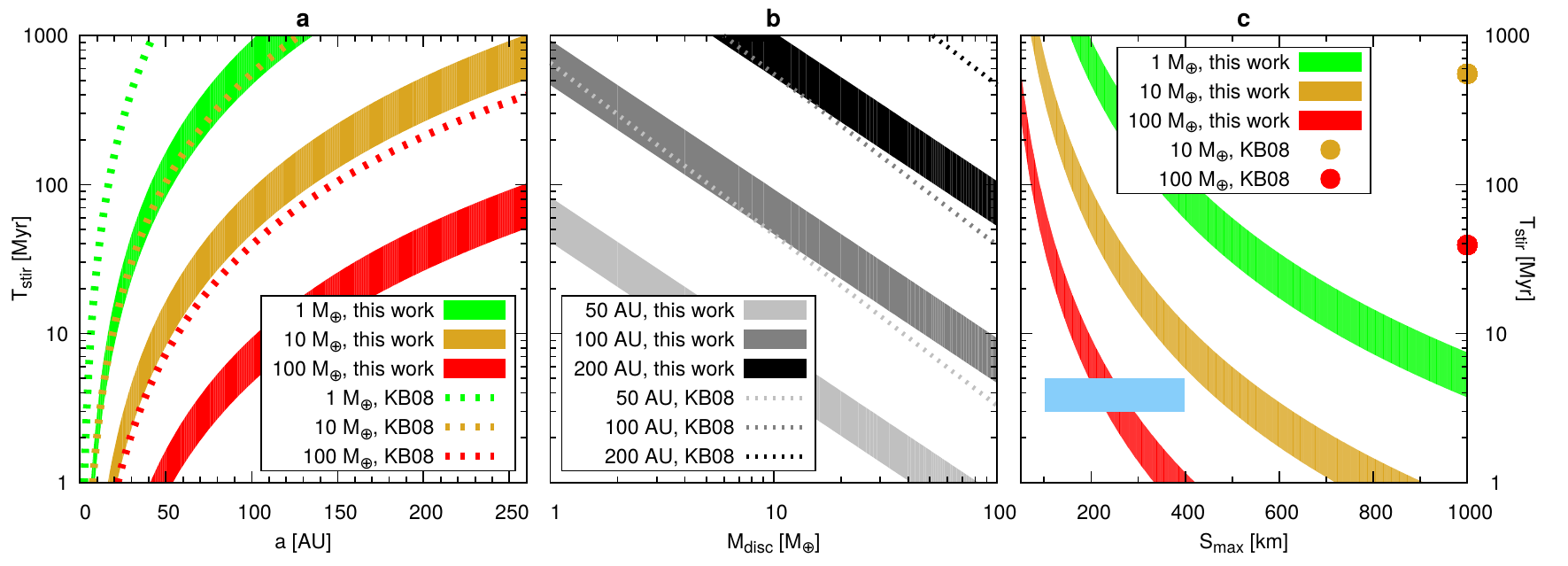}
\caption{
Stirring timescales calculated with Eqs.~(\ref{eq:Tstir parm}) and~(\ref{eq:1000 timescale parm}).
Nominal case is
$\rho = 1.0 \g\cm^{-3}$,
$v_\text{frag} = 30\m\s^{-1}$,
$S_\text{max} = 200\km$,
$\alpha = 1.6$,
$M_\star = M_\odot$,
$a = 100\AU$,
$\Delta a = 10\AU$,
and $M_\text{disc} = 100 M_\oplus$.
Panels show the effect of changing the following parameters:
(a) disc radius $a$ for a few values of disc mass $M_\text{disc}$,
(b) disc mass $M_\text{disc}$ for a few values of disc radius $a$,
and (c) radius of the largest planetesimals $S_\text{max}$ for a few values of disc mass $M_\text{disc}$.
Each filled area spans $\gamma$ from 1 (upper limit on the timescale) to 2 (lower limit).
The Kenyon \& Bromley timescales $T_\text{1000}$ for formation of $1000\km$-sized bodies
are shown with dotted lines in panels (a) and (b)
and with filled circles in panel (c).
Light-blue bar in panel (c) depicts the typical timescale
$T_\text{form} \approx (3\ldots 5)\Myr$
of planetesimal formation by pebble concentration.
The horizonal extent of that bar roughly reflects a conceivable range of $S_\text{max}$.
\label{fig:timescales}
   }
\end{figure*}

For the surface number density of planetesimals, we assume a power law
\be
  n = C \left( M/M_\text{max} \right)^{-\alpha},
\label{eq:n(M)}
\ee
where the normalisation factor $C$ can be determined from the total ring mass, $M_\text{disc}$.
Using
\be
 M_\text{disc} =
 2\pi a \; \Delta a \;
 \int_m^{M_\text{max}} M n(M)dM
\label{eq:M_disc}
\ee
results in
\be
 C \approx
 { M_\text{disc} \over M_\text{max}^2}
 {2 - \alpha \over  2\pi a \; \Delta a } .
\ee
Evaluating the integral in Eq.~(\ref{eq:1/T cont orig}) gives
\be
  T^{-1} =
  {1 \over 2 \pi}
  C_e \Omega
  \left( a \over \Delta  a\right)
  \left( M_\text{max} \over M_\star \right)
  \left( M_\text{disc} \over M_\star \right)
  {2 - \alpha \over 3 - \alpha} ,
\label{eq:1/T cont}
\ee
which is independent of $m$ and only differs from 
Eq.~(\ref{eq:1/T}) by the last term containing the power law index $\alpha$.

The solution to Eq.~(\ref{eq:dx/dt cont simple}) is
\be
 \sqrt{\left<e^2\right>}
 =
 \left( 2 \gamma \; t/T \right)^{1/4},
\label{e(t) cont}
\ee
and Eq.~(\ref{eq:Tstir efrag}) for the stirring timescale is replaced by
\be
  T_\text{stir} =  {T \over 2\gamma} e_\text{frag}^4 .
\label{eq:Tstir efrag cont}
\ee
Eqs.~(\ref{eq:n(M)}) and (\ref{eq:M_disc}) show that for $\alpha \le 2$
stirring comes from the largest planetesimals, i.e., those of mass close
to $M_\text{max}$.

\section{Stirring timescales}

\subsection{Timescales expressed through disc mass}

The stirring timescale~(\ref{eq:Tstir efrag cont}) can be reformulated
to show its dependence on all the model parameters:
\bea
  T_\text{stir}
  \!\!
  \!\!
  \!\!
  &=&
  \!\!
  \!\!
  \!\!
  9.3 \Myr
\nonumber\\
  &\times&
  \!\!
  \!\!
  \!\!
  \!\!
  \!\!
  \left( 1 \over \gamma \right)
  \left( \rho \over 1\g\cm^{-3} \right)^{-1}
  \left( v_\text{frag} \over 30\m\s^{-1} \right)^4
  \left( S_\text{max} \over 200\km \right)^{-3}
\nonumber\\
  &\times&
  \!\!
  \!\!
  \!\!
  \!\!
  \!\!
  \left( M_\star \over M_\odot \right)^{-1/2}
  \left( a \over 100\AU \right)^{7/2}
  \left( \Delta a /a  \over 0.1 \right)
  \left( M_\text{disc} \over 100 M_\oplus \right)^{-1} 
  \!\!
  \!\!
  \!\!
,
\label{eq:Tstir parm}
\eea
where we assumed $\alpha = 1.6$. 
This is justified by the fact that the models by
\citet{johansen-et-al-2015} and \citet{simon-et-al-2016,simon-et-al-2017} independently and 
robustly predict $\alpha = 1.6 \pm 0.1$.
We also note that the stirring timescale depends on $\alpha$ only 
weakly, see Eq.~(\ref{eq:1/T cont}).

It is natural to compare these results to the \citet{kenyon-bromley-2008} timescales
(see their Eq. 41):
\be
 T_\text{stir}^\text{KB}
 \sim
 T_\text{1000}
 =
 145 \Myr \;
 x_\text{m}^{-1.15}
 \left(a \over 80\AU \right)^3
 \left(2 M_\odot \over M_\star \right)^{3/2} ,
\label{eq:1000 timescale detail}
\ee
where $x_\text{m}$ can be expressed through the debris ring mass, location, and width
(see their Eq. 27):
\be
 x_\text{m}
 =
 \left( \Sigma \over \Sigma_0 \right)
 \left( a \over a_0 \right)^{3/2}
\label{eq:x_mmsn}
\ee
with
$a_0 = 30\AU$,
$\Sigma_0 = 0.18 \g\cm^{-2}(M_\star/M_\odot)$,
and
$\Sigma = M_\text{disc} / (2 \pi a \Delta a)$.
These formulae can be brought to the same form as Eq.~(\ref{eq:Tstir parm}):
\bea
&&
 T_\text{stir}^\text{KB}
  =
  39 \Myr
\nonumber\\
&&
  \times
  \left( M_\star \over M_\odot \right)^{-0.35}
  \left( a \over 100\AU \right)^{3.575}
  \left( \Delta a /a  \over 0.1 \right)^{1.15}
  \left( M_\text{disc} \over 100 M_\oplus \right)^{-1.15} .
\label{eq:1000 timescale parm}
\eea

Some typical results obtained with Eq.~(\ref{eq:Tstir parm})
are shown in Fig.~\ref{fig:timescales}.
\citeauthor{kenyon-bromley-2008}'s timescales
given by Eq.~(\ref{eq:1000 timescale parm}) are overplotted for comparison.
As a caveat, 
\citet{kenyon-bromley-2008} find the above formulae as a fit to their results
for the range $x_\text{m} \in [1/3,3]$ only.
The models with $M_\text{disc} = 100 M_\oplus$ and $1 M_\oplus$
would correspond to $x_\text{m} = 13.8$ and $0.14$, respectively.
However, another case also shown in the figure, $M_\text{disc} = 10 M_\oplus$,
has $x_\text{m} = 1.4$, for which the \citet{kenyon-bromley-2008} model is valid.

Figure~\ref{fig:timescales} 
provides a justification to Eqs.~(\ref{eq:one timescale})
and~(\ref{eq:stir timescale}).
Indeed, for the pebble concentration scenario,
it shows that the planetesimal formation timescale is negligible compared
to the disc excitation timescale,
i.e., $T_\text{form} \ll T_\text{excite}$.
And conversely, for the slow growth scenario, formation of Pluto-sized
objects takes much longer than the disc excitation by Plutos after their formation,
i.e., $T_\text{form} \gg T_\text{excite}$.
To draw the latter conclusion, we use the fact that the excitation timescales
determined in this work are independent of the assumed planetesimal formation scenario
and so also apply to the slow growth model.

Most importantly, Fig.~\ref{fig:timescales} demonstrates that the disc stirring
by planetesimals of 100s kilometres in radius formed by pebble concentration occurs more rapidly
than stirring by Plutos grown in the classical accretion scenario.
At the same time, we see that the timescales we predict are still long enough
to be taken into account.

\subsection{Timescales expressed through $x_\text{m}$}

\begin{figure*}
\centering
% Original at /astro/krivov/SMALL_PROJECTS/SELF_STIRRING/ANAL/timescales_MMEN.pdf
\includegraphics[width=1.0\textwidth, angle=0]{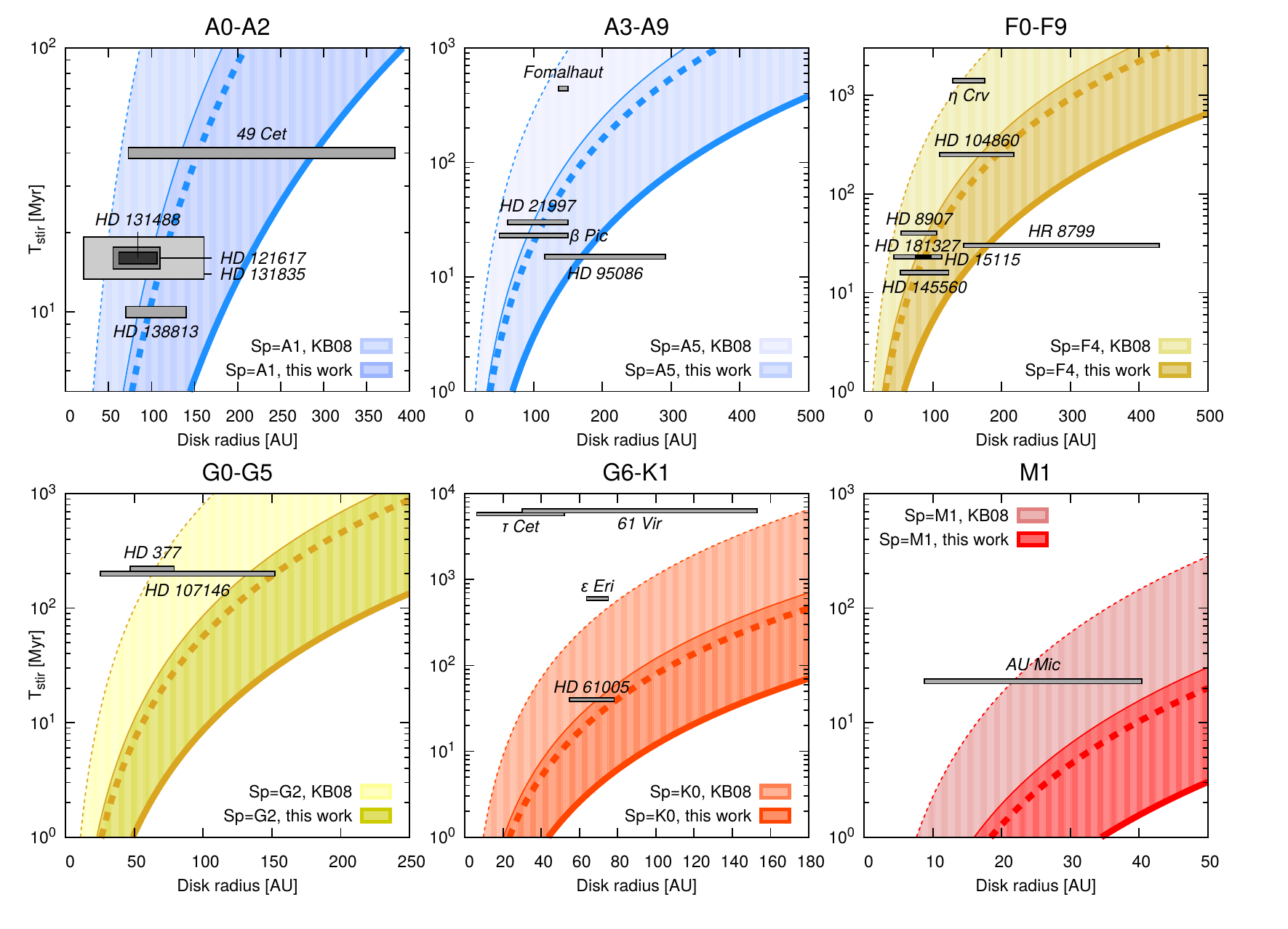}
\caption{Expected stirring timescales (filled areas with lines) in comparison with selected prominent debris discs
(bars showing the radial disc extent and having an arbitrary vertical thickness).
Darker filling colours bordered by solid lines: our Eq.~(\ref{eq:Tstir MMEN}) with  $\gamma = 1.5$,
lighter filling colours bordered by dashed lines: Kenyon \& Bromley's Eq.~(\ref{eq:1000 timescale MMEN}).
Thin and thick lines correspond to density scalings $x_\text{m}$ of 1 and 10, respectively.
The other parameters are fixed:
$\rho = 1.0 \g\cm^{-3}$,
$v_\text{frag} = 30\m\s^{-1}$,
$S_\text{max} = 200\km$,
and $\alpha = 1.6$.
Panels from top left to bottom right correspond to stellar type ranges:
A0--A2, A3--A9, F0--F9, G0--G5, G6--K1, and M1.
Colours just roughly mimic the photospheric colours for these spectral type ranges.
\label{fig:timescales MMEN}
   }
\end{figure*}

Instead of taking $M_\text{disc}$ as one of the parameters, 
we may use \citeauthor{kenyon-bromley-2008}'s parameter $x_\text{m}$.
This parameter is a measure of the disc mass in the units of MMSN.
We use Eq.~(\ref{eq:x_mmsn}) to find
\be
  \left( M_\text{disc} \over 100 M_\oplus \right)
  =
  0.0726 x_\text{m}
  \left( M_\star \over M_\odot \right)
  \left( a \over 100\AU \right)^{1/2}
  \left( \Delta a /a  \over 0.1 \right).
\ee
Then,
\bea
  T_\text{stir}
  \!\!
  \!\!
  \!\!
  &=&
  \!\!
  \!\!
  \!\!
  {129 \Myr \over x_\text{m}}
\nonumber\\
  &\times&
  \!\!
  \!\!
  \!\!
  \!\!
  \!\!
  \left( 1 \over \gamma \right)
  \left( \rho \over 1\g\cm^{-3} \right)^{-1}
  \left( v_\text{frag} \over 30\m\s^{-1} \right)^4
  \left( S_\text{max} \over 200\km \right)^{-3}
\nonumber\\
  &\times&
  \!\!
  \!\!
  \!\!
  \!\!
  \!\!
  \left( M_\star \over M_\odot \right)^{-3/2}
  \left( a \over 100\AU \right)^3
  \!\!
  \!\!
  \!\!
\label{eq:Tstir MMEN}
\eea
and
\be
 T_\text{stir}^\text{KB}
  =
  {801 \Myr \over x_\text{m}^{1.15}}
  \left( M_\star \over M_\odot \right)^{-3/2}
  \left( a \over 100\AU \right)^3 .
\label{eq:1000 timescale MMEN}
\ee

Unlike Eqs.~(\ref{eq:Tstir parm})
and~(\ref{eq:1000 timescale parm}),
Eqs.~(\ref{eq:Tstir MMEN})
and~(\ref{eq:1000 timescale MMEN})
are independent of the debris ring width $\Delta a$,
and the dependence on other parameters is different.
For instance, 
Eqs.~(\ref{eq:Tstir MMEN})--~(\ref{eq:1000 timescale MMEN})
predict a stronger dependence on the stellar mass
than Eqs.~(\ref{eq:Tstir parm})
and~(\ref{eq:1000 timescale parm}).
This is because the parameter $x_\text{m}$, as defined by
\citet{kenyon-bromley-2008}, includes $M_\star$.
A physical basis for their definition is that the protoplanetary
disc masses are known to be roughly proportional to the stellar masses
\citep[e.g.,][]{williams-cieza-2011}, so that an ``MMSN'' of a lower-mass star
has a lower mass than the one around a higher-mass star.

\section{Applications}

We now apply the results to a handful of prominent debris discs to see
which of them can and which cannot be self-stirred over their full radial extent.
We took a sample of discs resolved by ALMA or SMA from \citet{matra-et-al-2018}.
Marginally resolved discs (marked with an asterisk in their Table~1) were excluded.
The resulting list includes 23 discs with a broad coverage of radii and ages
around stars from early to late types.

Figure~\ref{fig:timescales MMEN} compares the ages and radial extent of the discs
in this sample with the predictions of the self-stirring models,
both Kenyon \& Bromley's and the one developed in this work.
It demonstrates that all discs can be classified into three groups:
\begin{enumerate}
\item
Some discs (including the majority of discs around stars later than G0)
can be stirred in both scenarios
(e.g., Fomalhaut, $\eta$~Crv, HD~377, 61~Vir, $\tau$~Cet, $\varepsilon$~Eri, AU~Mic).
\item
For some others, however, Kenyon \& Bromley's scenario fails, whereas stirring by
$200\km$-sized planetesimals would work
(e.g., HD~131835, HD~138813, $\beta$~Pic, HD~145560).
\item
Still others do require planets (49~Cet, HD~95086, HR~8799).
At least the outer parts of their extended planetesimal belts cannot be self-stirred for any 
reasonable parameter choices in our model. 
\end{enumerate}

The three systems that require planets to explain why their debris discs are stirred are all 
well-known and truly remarkable:
\begin{itemize}
\item
One is HR~8799,
a young ($\sim 30\Myr$) system with a planetesimal belt extending from $\approx 140\AU$ to as far as
$\approx 440\AU$ from the star \citep{booth-et-al-2016}. In this system, four planets have been discovered
by direct imaging \citep{marois-et-al-2008,marois-et-al-2010}, 
and one more has been suggested \citep{booth-et-al-2016,read-et-al-2018}
(see, however, \citeauthor{wilner-et-al-2018} \citeyear{wilner-et-al-2018}).
\item
Another one is HD~95086, a $17 \pm 4 \Myr$-old \citep{moor-et-al-2013} star with a broad disc extending from
$106 \pm 6 \AU$ to $320 \pm 20 \AU$ \citep{su-et-al-2017b,zapata-et-al-2018} and one
massive planet \citep{rameau-et-al-2013,rameau-et-al-2014}.
\item
The third system is 49~Cet, a $40\Myr$-old \citep{zuckerman-song-2012} star with a large and broad disc
\citep{hughes-et-al-2017,choquet-et-al-2017}.
One peculiarity of this system is that it is currently the oldest one found to harbour molecular 
gas in copious amounts \citep[see., e.g.,][]{kral-et-al-2017c}.
No planets have been discovered so far around 49~Cet, however.
\end{itemize}

Since 49~Cet is the only system in this group without known planets, it deserves a closer look.
Using the \citet{mustill-wyatt-2009} planet stirring formulae,
\citet{moor-et-al-2014} made calculations to estimate the parameters of
an alleged planet that would stir the entire 49~Cet disc.
At the time only low-resolution {\em Herschel} data were available and so 
they used a conservative estimate of $250\AU$ as the outer edge of the planetesimal belt.
From these calculations they found that a low eccentricity ($\la 0.05$) planet
could only stir out to such a distance
if it were massive ($>6 M_\text{jup}$) and close to the inner edge of the disc at $70\AU$
(see their Fig.~7).
Such high mass planets are clearly ruled out by the observational limits of SPHERE observations that 
range from $\sim 3 M_\text{jup}$ at $20\AU$ to $\sim 1 M_\text{jup}$ at $110\AU$ \citep{choquet-et-al-2017}.
For a planet to satisfy these limits, it would need an eccentricity of at least 0.2 to stir out
to $250\AU$ and higher to stir out to the full extent 
of the disc as now seen by ALMA \citep[$>300\AU$,][]{hughes-et-al-2017}. 
It is possible though that lower planetary masses would suffice if two or more planets were
present in the system \citep[e.g.,][]{lazzoni-et-al-2018}.

Following \citet{matra-et-al-2018}, the above has assumed a simplistic model
of a smooth disc with a sharp inner and outer edge.
\citet{hughes-et-al-2017} note an alternative possibility suggested by the observations
of a narrow ring located at $110\AU$ combined with a broad disc,
where the emission beyond $110\AU$ is coming from small (possibly primordial) grains and 
so the disc only needs to be stirred out to the location of the ring
(see also a discussion in \citeauthor{krivov-et-al-2013} \citeyear{krivov-et-al-2013}).
Assuming this is the case, they then show that a planet 
responsible for this could easily have a mass lower than the observational limits,
whilst also having a low eccentricity orbit.
However, if the disc only needs to be stirred out to $110\AU$,
our results show that self-stirring can explain the dust in this system
as the self-stirred region of our model extends out to $\sim 250\AU$ for an age of 40 Myr. 
Nonetheless, further work is necessary to determine the exact source
of these grains at large distance and whether they really can be explained
without the need for a collisional cascade at large distances.

Coincidentally or not, all three discs in our sample that require planets are those around stars
with spectral classes earlier than F0.
This may be surprising, as the stirring timescale is shorter around more massive stars
(see  Eq.~\ref{eq:Tstir MMEN}).
However, this is probably a double bias in our disc selection.
First, early-type stars are younger on the average,
and second, their discs are larger on the average \citep{matra-et-al-2018}.
Obviously, younger and/or larger discs are more difficult to explain by self-stirring.

It is also interesting to compare our analysis with that of \citet{moor-et-al-2014}.
They addressed the same question of whether the discs can or cannot be self-stirred,
based on a somewhat different sample which included 11 bright discs well-resolved by
{\em Herschel}. Five of them (49~Cet, HD~21997, $\beta$~Pic, HD~95086, and HR~8799)
are also part of our sample. Their Fig.~6 compares the radii and age estimates of the discs
(not exactly the same as adopted here) with the predictions of
\citet{kenyon-bromley-2008} model.
For three out of five discs that appear in both samples,
our conclusion is the same as theirs: 
49~Cet, HD~95086, and HR~8799 cannot be self-stirred.
For the other two discs (HD~21997 and $\beta$~Pic), the conclusion is different: these discs
cannot be self-stirred according to their study, but are compatible with being self-stirred
in this work. This difference is readily understood by inspecting Fig.~\ref{fig:timescales MMEN}.
It shows that the ages of these two systems are younger than the timescale of disc 
stirring by Plutos, but older than the stirring timescale by smaller planetesimals formed
by pebble concentration.  
There is one more disc in their sample, HD~16743 with an estimated radius of $157 \pm 20\AU$
and an age of $10$--$50\Myr$, that appears incompatible with self-stirring.
Assuming this radius and age, that disc could be self-stirred in our model.

We can also apply our self-stirring models to other samples. For instance,
\citet{holland-et-al-2017} reported on outer radii of 16 debris discs resolved in the SONS survey
done with the JCMT/SCUBA-2 sub-mm camera (see their Tab.~4).
Seven of them are not part of our sample.
We computed the self-stirring timescales for these discs by means of 
Eq.~(\ref{eq:Tstir MMEN}) with the same set of parameters as for our sample.
We then compared them with the system ages listed in Tab. 1 of \citet{holland-et-al-2017}.
Five of the discs
(HD~38858, HD~48682, $\gamma$~Oph, HD~170773, Vega, and HD~207129)
turned out to be compatible with self-stirring.
Two other discs (HD~15745 and HD~143894) would require planets as stirrers,
at least to explain them over the full radial extent up to the outer edge.
These conclusions come with the caveat that the angular resolution of SCUBA2
is lower than that of ALMA and SMA, so that the disc radii given in 
\citet{holland-et-al-2017} are more uncertain than the ones in our sample.

\section{Discussion}

\subsection{Uncertainties of the model}

We argue that our model is likely setting an upper limit on the self-stirring timescales, which is to say
that in reality the cascade may ignite earlier.
There are several reasons to expect this.
Firstly, we only considered self-stirring by planetesimals
that form ``instantaneously'' by particle concentration models.  Our model does not take into account 
that these planetesimals, whose sizes are originally smaller than a few 100s of kilometres, may~-- and 
most likely will~-- grow further to Pluto and even gas giant core sizes, either in 
traditional Kenyon--Bromley-Kobayashi mode or by other mechanisms such as pebble accretion
\citep{lambrechts-johansen-2012,lambrechts-johansen-2014}.
Should that subsequent growth proceed faster than the stirring 
timescales considered here, the cascade will obviously ignite earlier.

Secondly, the stirring timescale drops rapidly with increasing size of the largest planetesimals.
Figure~\ref{fig:timescales MMEN} assumes $S_\text{max} = 200\km$. If the largest bodies are somewhat larger 
than $200\km$, for instance $300\km$, which cannot be excluded \citep{johansen-et-al-2015,simon-et-al-2016,
schaefer-et-al-2017,simon-et-al-2017}, the stirring timescale will shorten by more than a factor of three.
Furthermore, the maximum planetesimal size may not be independent of some of the other factors
in Eq.~(\ref{eq:Tstir parm}).
For instance, $S_\text{max} \propto a^{3/2}$ \citep{schaefer-et-al-2017}, meaning that larger 
planetesimals form farther out from the star.
Taking this into account would flatten the dependence of $T_\text{stir}$ on $a$, speeding up the
stirring in the outer parts of the discs.

Apart from $S_\text{max}$, there are a few more poorly known parameters in our model.
One is the bulk density $\rho$, which we set to $1\g\cm^{-3}$ in 
Fig.~\ref{fig:timescales MMEN}. Using, for instance,
$0.53\g\cm^{-3}$ instead, as inferred for the comet 67P/Churyumov-Gerasimenko
\citep{jorda-et-al-2016}, would double the stirring timescale.

However, most of the uncertainty probably comes from the minimum velocity 
for fragmentation, $v_\text{frag}$.
Our choice in the numerical examples above, $30\m\sec^{-1}$, is rather arbitrary.
This velocity is directly related to (actually, is roughly a square root of) the critical 
fragmentation energy $Q_\text{D}^*$, determination of which has been the subject
of numerous laboratory impact experiments and hydrocode simulations 
\citep[see, e.g.,][]
{blum-wurm-2008,stewart-leinhardt-2009,guettler-et-al-2010,leinhardt-stewart-2012,blum-2018}.
The critical fragmentation energy depends on the composition of planetesimals, being quite
different for the ``pebble piles'' predicted by turbulent concentration and streaming 
instability models and for the ``monolithic'' bodies formed in traditional slow growth models
\citep[see, e.g.,][and references therein]{krivov-et-al-2018}.
Furthermore, the composition and strength may be different at different distances from the star, 
and may even change in time during early evolution of the discs.
Of course, $Q_\text{D}^*$ and so $v_\text{frag}$
are also strong functions of size. Sizes that matter for the stirring calculations are those
for which collisional timescales do not exceed the current age of the systems. Thus these vary 
in time and depend on the distance from the star as well.
Another complication arises from the fact that some degree of fragmentation is possible even 
at velocities insuficient for collisional disruption, through erosive collisions
\citep[e.g.,][]{kobayashi-et-al-2010b}.
All this makes choosing the right value for $v_\text{frag}$ very difficult.
At the same time, the stirring timescale depends on it very 
sensitively, since $T_\text{stir} \propto v_\text{frag}^4$. Overall, the resulting timescales
we infer can easily be uncertain to at least one order of magnitude, perhaps even more.

One more parameter that strongly affects the predicted stirring timescales is the
total mass of a planetesimal disc.
In all the examples given in this paper, we do not consider
planetesimal discs with mass larger than $100 M_\oplus$, because this
would exceed the total mass of solids in protoplanetary progenitors to debris discs,
inferred from \hbox{(sub-)mm} surveys \citep[e.g.,][]{williams-cieza-2011}.
However, \citet{krivov-et-al-2018} have shown that bright discs (such as those 
considered here) require higher total masses in planetesimals, up to several $1000 M_\oplus$,
to be explained with steady-state collisional models. One possibility is that protoplanetary discs are indeed
more massive ($\sim 0.1$-- a few $M_\star$) and larger ($\sim 100$--$1000\AU$)
than usually assumed \citep{nixon-et-al-2018}.
Such discs would obviously become gravitationally unstable and might be able to build planetesimals pretty early.
In this case, the total (unobservable) mass of planetesimal rings in the outer systems might, indeed, be much
higher than $\sim 100 M_\oplus$. This would dramatically shorten self-stirring timescales, meaning that
even the young and large discs do not necessarily require planets~-- which does not, however,
exclude the possibility that these are present, as such massive discs should also form planets quickly.

\subsection{Implications of the model}

Notwithstanding the uncertainties, we find that sufficiently young and large discs
are clearly incompatible with self-stirring.
Assuming that discs can only be stirred either by embedded planetesimals
or by planets, this automatically means that one or more planets must be present in such systems.
Furthermore, the stirring criterion is actually a more compelling diagnostic of planets than the other
ones commonly invoked. One of those is the fact that all of the debris discs have inner cavities.
Even though these are often attributed to planets inside the discs
\citep[e.g.,][]{shannon-et-al-2016,zheng-et-al-2017,lazzoni-et-al-2018,regaly-et-al-2017},
planetesimals might preferentially form in distinct radial zones 
\citep[e.g.,][]{carrera-et-al-2017}.
Similarly, many of the discs exhibit asymmetries that are also considered signposts of planets
\citep[e.g.,][]{lee-chiang-2016,loehne-et-al-2017}.
Yet here, too, there exist alternative scenarios to explain the asymmetries.
These include interactions with the ambient interstellar medium \citep{debes-et-al-2009},
recent giant impacts \citep{olofsson-et-al-2016},
gravitational interactions in a debris disc-hosting multiple stellar system \citep{shannon-et-al-2014,kaib-et-al-2017},
and combinations of these effects \citep[e.g.][]{marzari-2012}.
While accounting for inner gaps or asymmetries, not all of these alternative models and scenarios explain {\it 
per se} why the planetesimal discs get stirred.
Thus it is the stirring requirement that points to planets more unambiguously.
A caveat is that stirring mechanisms other than self-stirring and planetary stirring cannot be completely excluded.
Some of the discs could be excited for instance by stellar flybys in birth cluster environments 
\citep{kenyon-bromley-2002,kobayashi-ida-2001} of by as yet undiscovered external stellar companions
\citep[e.g.,][]{thebault-et-al-2010,thebault-2012}.

\section{Conclusions}

In this paper, we investigate the birth stage of debris discs.
Since the planetesimals left over after the gas dissipation should be in
low-eccentricity, low-inclination orbits, some mechanism is required
to dynamically excite them to relative velocities above the fragmentation threshold,
allowing them to produce observable debris dust in mutual collisions.
One natural mechanism would be ``self-stirring'', in which smaller planetesimals are
excited by larger planetesimals embedded in the disc. Following \citet{kenyon-bromley-2008},
it is commonly assumed that self-stirring comes into play as soon as the first Pluto-sized
objects have formed. We explore the idea that smaller objects, with radii of $\la 200\km$,
that are predicted to form in the disc through pebble concentration by the time of gas dispersal,
may be able to excite the planetesimal disc well before the first Plutos are able to form.
Our main conclusions are as follows:
\begin{itemize}
\item
We conclude that $1000\km$-sized objects are, indeed, not really necessary to stir debris discs.
Planetesimals $\sim 200\km$ in size are sufficient.
Although $1000\km$-sized objects would stir a disc promptly, their formation takes longer
than the time it takes for much more rapidly forming $200\km$-sized objects to stir the same disc.
\item
Applying the model to a suite of debris discs resolved in the sub-mm,
we show that the majority of them could be
self-stirred by $\sim 200\km$-sized planetesimals.
However, we have identified three discs (HR~8799, HD~95086, and 49~Cet) that
cannot be explained by self-stirring.
Such systems would be the natural targets to search for planets.
Indeed, we note that planets are already known around two of these.
Further resolved observations of young discs will likely produce more candidates for discs where 
planet stirring is necessary and so more promising targets for planet searches.
\end{itemize}

\section*{acknowledgments}

We thank the reviewer for a number of useful and constructive comments
and Scott Kenyon for enlightening discussions.
This research has been supported by the {\it Deutsche Forschungsgemeinschaft (DFG)}
through grants Kr~2164/13-1 and Kr~2164/15-1.

%%%%%%%%%%%%%%%%%%%%%%%%%%%%%%%%%%%%%%%%%%%%%%%%%%


\newcommand{\AAp}      {Astron. Astrophys.}
\newcommand{\AApR}     {Astron. Astrophys. Rev.}
\newcommand{\AApS}    {AApS}
\newcommand{\AApSS}    {AApSS}
\newcommand{\AApT}     {Astron. Astrophys. Trans.}
\newcommand{\AdvSR}    {Adv. Space Res.}
\newcommand{\AJ}       {Astron. J.}
\newcommand{\AN}       {Astron. Nachr.}
\newcommand{\AO}       {App. Optics}
\newcommand{\ApJ}      {Astrophys. J.}
\newcommand{\ApJL}      {Astrophys. J. Lett.}
\newcommand{\ApJS}     {Astrophys. J. Suppl.}
\newcommand{\ApSS}     {Astrophys. Space Sci.}
\newcommand{\ARAA}     {Ann. Rev. Astron. Astrophys.}
\newcommand{\ARevEPS}  {Ann. Rev. Earth Planet. Sci.}
\newcommand{\BAAS}     {BAAS}
\newcommand{\CelMech}  {Celest. Mech. Dynam. Astron.}
\newcommand{\EMP}      {Earth, Moon and Planets}
\newcommand{\EPS}      {Earth, Planets and Space}
\newcommand{\GRL}      {Geophys. Res. Lett.}
\newcommand{\JGR}      {J. Geophys. Res.}
\newcommand{\JOSAA}    {J. Opt. Soc. Am. A}
\newcommand{\MemSAI}   {Mem. Societa Astronomica Italiana}
\newcommand{\MNRAS}    {MNRAS}
\newcommand{\PASJ}     {PASJ}
\newcommand{\PASP}     {PASP}
\newcommand{\PSS}      {Planet. Space Sci.}
\newcommand{\QJRAS}    {Quart. J. Roy. Astron. Soc.}
\newcommand{\RAA}      {Research in Astron. Astrophys.}
\newcommand{\SolPhys}  {Sol. Phys.}
\newcommand{\SolSysRes}{Sol. Sys. Res.}
\newcommand{\SSR}      {Space Sci. Rev.}

\input ms.bbl

%%%%%%%%%%%%%%%%%%%%%%%%%%%%%%%%%%%%%%%%%%%%%%%%%%
% Don't change these lines
\bsp    % typesetting comment
\label{lastpage}
\end{document}